# A fairness assessment of mobility-based COVID-19 case prediction models

Abdolmajid Erfani [1*], and Vanessa Frias-Martinez [2, 3]

[1] Department of Civil and Environmental Engineering, University of Maryland, 1173 Glenn Martin Hall, College Park, MD 20742, USA.

[2] College of Information Studies, University of Maryland, College Park, MD 20742, USA.

[3] University of Maryland Institute for Advanced Computer Studies, University of Maryland, College Park, MD 20742, USA

* Corresponding Author: erfani@umd.edu

# ABSTRACT

In light of the outbreak of COVID-19, analyzing and measuring human mobility has become increasingly important. A wide range of studies have explored spatiotemporal trends over time, examined associations with other variables, evaluated non-pharmacologic interventions (NPIs), and predicted or simulated COVID-19 spread using mobility data. Despite the benefits of publicly available mobility data, a key question remains unanswered: are models using mobility data performing equitably across demographic groups? We hypothesize that bias in the mobility data used to train the predictive models might lead to unfairly less accurate predictions for certain demographic groups. To test our hypothesis, we applied two mobility-based COVID infection prediction models at the county level in the United States using SafeGraph data, and correlated model performance with sociodemographic traits. Findings revealed that there is a systematic bias in models' performance toward certain demographic characteristics. Specifically, the models tend to favor large, highly educated, wealthy, young, and urban counties. We hypothesize that the mobility data currently used by many predictive models tends to capture less information about older, poorer, less educated and people from rural regions, which in turn negatively impacts the accuracy of the COVID-19 prediction in these areas. Ultimately, this study points to the need of improved data collection and sampling approaches that allow for an accurate representation of the mobility patterns across demographic groups.

# INTRODUCTION

The interactions between human mobility and epidemic spread have been studied unprecedentedly during the COVID-19 pandemic [1-8]. With these efforts, nonpharmaceutical interventions (such as national lockdowns) have been evaluated for their effectiveness and socio-economic impact on different groups [9-11], models have been developed to predict disease spatial diffusion [12,13], and scenarios have been modeled to assess their outcomes [14-17]. Studies have demonstrated that mobility data are a meaningful proxy measure of social distancing [18], affect viral spreading [19,20], and are useful for predicting the spread of COVID-19 [21-23].

In particular, to control the spread of new cases and plan efficiently for hospital needs and capacities during an epidemic, public health decision-makers require accurate predictions of future case numbers [7]. For example, a study by Ilin et al. (2021) showed that changes in mobility can be used to predict COVID-19 cases. Their study demonstrated that public mobility data can be used to develop reduced-form and simple models that mimic the behavior of more sophisticated epidemiological models for predicting COVID-19 cases on a 10-day basis [21]. Another study examined several state-of-the-art machine learning models and statistical methods and demonstrated how mobility data can improve prediction trends when used as exogenous information in models [22].

As discussed, mobility data from anonymized smartphones has been shown to improve COVID-19 case prediction models. However, mobility data bias has received little attention in this predictive context. There exist only just a handful of papers reporting demographic bias in mobility data due to differences in smartphone ownership and use [24-26]; and since data providers are not transparent about how mobility data is collected, or about the socio-economic and demographic groups represented in them, directly measuring and correcting bias in mobility data is difficult. In this study, we hypothesize that the presence of socio-economic and demographic bias in the mobility data used to train the COVID-19 case predictive models, might result in unfairly less accurate predictions for particular socio-economic and demographic groups. Unfair predictions provided to decision makers e.g., predictions of COVID-19 cases for minority groups that are lower than reality, could in turn be used to unfairly assign more resources to population groups that do not necessarily need them.

To test our hypothesis, we evaluated the performance of two types of mobility-based COVID-19 case prediction models highly used by decision makers due to its interpretability: linear regressions and time series models. In contrast to more complex epidemiological models that are hard to tune due to its parametric nature, and deep learning models that are difficult to interpret, linear models and time series are easy to train and test [21,27-29]. The models were trained using SafeGraph's mobility data, and performance was measured via predictive errors. To assess the fairness of the predictions, we analyzed the relationship between the model prediction errors and specific socio-economic and demographic features at the county level in the United States and across the two model types. Evaluating the performance of two diverse interpretable models allowed us to account for potential algorithmic bias i.e., bias introduced by the algorithm itself [30,31]. If unfair predictions are pervasive across types of models trained and tested with the same data, we can partially attribute the unfairness to the mobility data itself.

# MATERIAL AND METHODS

In our study, we use mobility data from SafeGraph to build COVID-19 case prediction models; and we explore model performance across socio-economic and demographic features to potentially identify unfair results for specific groups i.e., differences in error distributions across social groups. We next describe these three types of datasets, with all being publicly available.

**Human mobility**

We used SafeGraph's publicly-available human mobility data at the county level in the US. SafeGraph uses location information extracted from smartphones to provide aggregate data characterizing mobility in terms of visit volumes to types of places and volumes of origin-destination (OD) flows [32]. For this study specifically, we used the data publicly available in the origin-destination-time (ODT) platform [33], that computes OD flows between counties as the aggregation of trips that start at an individual's home county location (origin), with a destination defined as a stay location within a county for longer than a minute. OD flows between all counties in the US were collected throughout all days of the year 2020. Figure 1.a illustrates how the average number of trips at county level across the US changed over the year 2020. According to various studies in the US using mobility data, the dataset collected in Figure 1.a also shows similar trends of mobility change [34,35].

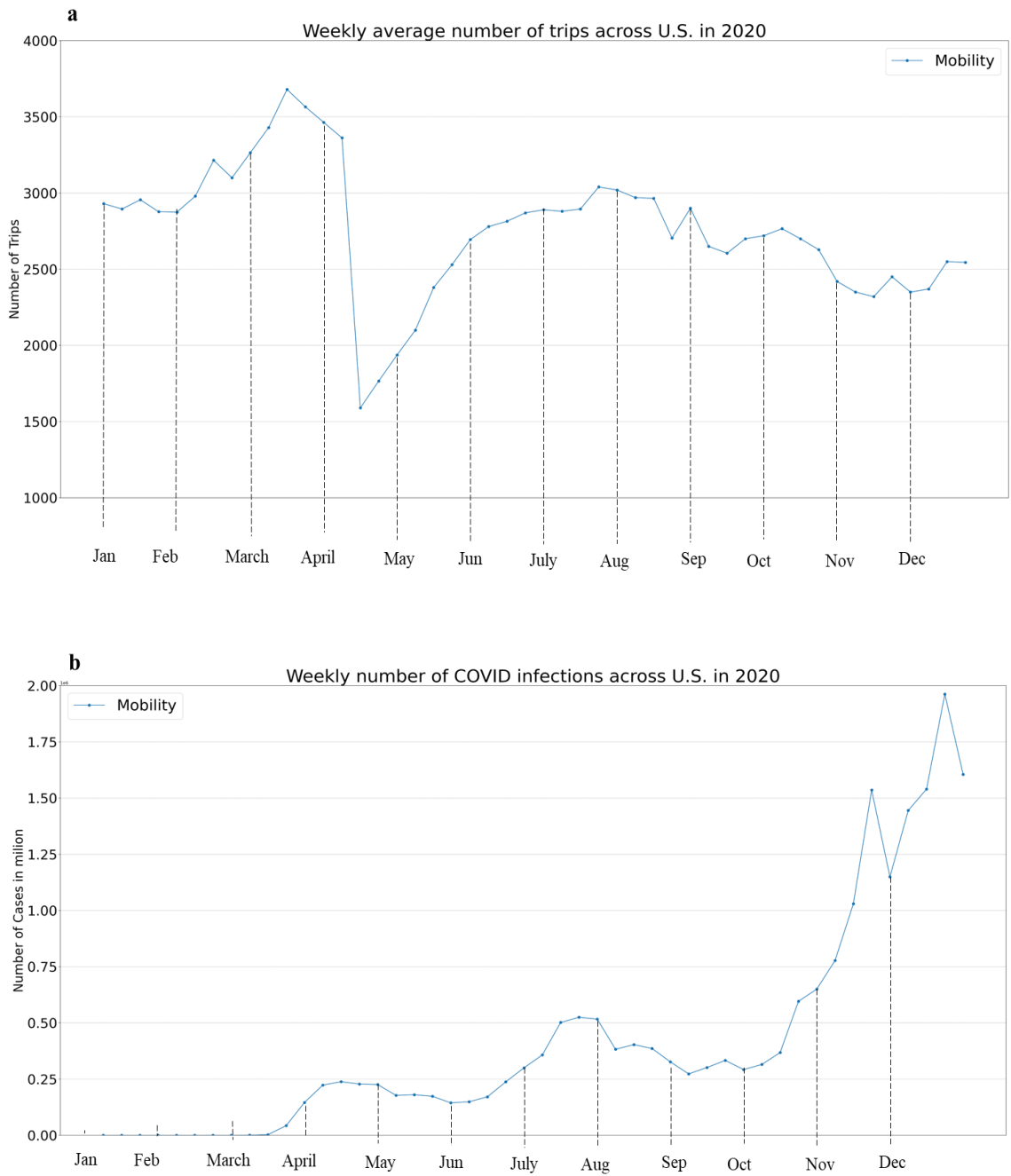

Figure 1. Data on mobility measures, COVID-19 infections. (a) Weekly average number of trips across the U.S. (b) Weekly new number of COVID infections across the U.S.

**COVID‑19 cases**

In order to obtain the cumulative and daily confirmed cases of COVID-19 for each county unit, we refer to the data repository compiled by the Johns Hopkins Center for Systems Science and Engineering (CSSE) [36]. As shown in Figure 1.b, the weekly number of new COVID infections has been increasing over the year 2020.

**Socio-economic and demographic data**

Data on socio-economic and demographic characteristics at the county level was also collected from public databases [37]. Studies have shown that sociodemographic factors such as age, race, income, educational level, and area of residence can influence smartphone ownership and usage, which may have an impact on mobility data biases [38,39]. Therefore, we collected a wide range of information, including the population, income, education, age, and ownership of smartphones. Also, we used the US National Center for Health Statistics Urban-Rural Classification Scheme for Counties [40], which assigns each county an ordinal code ranging from 1 (most urban) to 6 (most rural).

# Methods

In this section we will describe (i) the two types of models used in the COVID-19 case prediction; (ii) the training and evaluation of these models; and (iii) the process proposed to evaluate the fairness of the predictions across socio-economic and demographic groups, as well as across models.

**Models**

Model 1: Linear regression (Ilin et al. [21], Wang et al. [41], Ayan et al. [42], and Sahin [43]). Several papers have suggested that linear regressions that combine mobility data with historical COVID-19 cases can successfully predict future cases [21,41-43]. These models generally use different lags between mobility rates and COVID-19 cases to account for the infection period i.e., the period between the person's movement – and potentially interaction with others and infection – and the person testing positive for COVID-19. For this study we use Ilin at al. linear model [21] because rather than picking one lag, they propose to consider multiple lags within the model encompassing the plethora of linear regressions that have been tested in the literature. Specifically, Ilin at al. (2021) use a distributed-lag model to estimate log confirmed infections as the dependent variable, with average mobility over lags 1-7, 8-14, and 15-21 days to predict the number of COVID-19 cases at a given day:

$$\log \frac{I_{it}}{I_{i,t-1}} = \beta_1 m_{1-7,it} + \beta_2 m_{8-14,it} + \beta_3 m_{15-21,it} + \epsilon_{it} \qquad (1)$$

where i is the unit of analysis, $\log \frac{I_{it}}{I_{i,t-1}}$ is the first difference of log confirmed cases at time t, and m represents mobility measures averaged over lags 1-7, 8-14 and 15-21, respectively.

Model 2: Time series forecasting (Aji et al. [28], Zhao et al. [29], Zeng et al. [44], and Klein et al. [45]). The Autoregressive Integrated Moving Average (ARIMA) model is a statistical method that considers both past and present data for forecasting. An ARIMAX model, also known as ARIMA with multiple regressors, extends the basic ARIMA model to include other external variables for prediction. In the COVID-19 setting, mobility data and other sources of information have been used as regressors to potentially improve the predictive models [29,44,45]. For example, in their study

Zhao et al. [29] conclude that with mobility data, time series forecasting provides accurate predictions with mobility data lags of between 8-10 days for dense or sparse populations respectively. In this study, we consider an ARIMAX (p, d, q) model that can be expressed as:

$$y_t = \beta_0 x_t + \sum_{j=1}^{p} \emptyset_j y_{t-j} + \varepsilon_t + \sum_{j=1}^{q} \theta_j \varepsilon_{t-j} \qquad (2)$$

where y is the number of confirmed infections, x is the mobility change as exogenous variable lagged by 21 days (similar to Model 1's selection of lag), p is the Autoregressive (AR) parameter, q is the Moving Average (MA) parameter, d is the degree of first differencing to make data stationarity, $\varepsilon$ is the error, and $\beta_0, \emptyset_j, \theta_j$ are model parameters to be esimated. By using the Python package Auto Arima, we were able to generate the best p, d, and q values based on the data set, thus providing better forecasts [46]. To summarize, the lag of mobility, historic number of COVID cases can be used to predict future cases at unit of analysis.

**Training and model evaluation**

To train the models, we used both historical COVID-19 data and mobility OD flows from mid-April to December 2020. Rather than using the raw mobility OD flows, we used a measure of mobility change over a baseline, which was calculated by dividing the daily mobility by the average daily mobility in February 2020, a non-holiday month before the COVID-19 pandemic. This is a common approach in prior COVID-19 case predictive models that use mobility data [21,29].

Different training lengths were evaluated for both models, and the ones with the best accuracies were selected. To train the linear regression (Model 1), we used a 21-day training set with a 1-day shifting window from April 14th to December 30th. The time series model (Model 2), was trained over a period of 90 days, resulting in predictions available from early August to the end of the year. Once trained, each model was used to predict the number of COVID-19 cases for two lookaheads: 1-day (next day) and 7-days time (week) intervals at the county level, as predictions on a daily and weekly basis are a common theme in previous studies [21,28,29,41-44]. In this process, thousands of regressions and ARIMAX models are trained at the county level on a daily basis to be able to predict 1 day and 7 days later how many COVID-19 cases will occur.

Finally, the model performance was evaluated via the error rate, which was calculated on a daily basis based on the difference between the actual number of COVID cases and predictions as Equation 3. A mean absolute percentage error rate (MAPE) is calculated by averaging the error rates for specific counties over a given time period.

$$Error\ rate_t = \left|\frac{Prediction\ value_t\ -\ Actual\ number\ of\ COVID\ cases_t}{Actual\ number\ of\ COVID\ cases_t}\right| \quad (3)$$

$$MAPE = \frac{100\%}{n}\sum_{t=1}^{n} Error\ rate_t \quad (4)$$

**Fairness Analysis.**

We analyzed the fairness of the predictions for each model by computing the weekly MAPE per lookahead (1-day and 7-day) at the county level, followed by a spearman rank-order correlation analysis between the average weekly error rate across counties in the US and their socio-economic and demographic characteristics presented in the data section: household income (average household income), smartphone ownership (percentage of households owning smartphones), population, education level (bachelor's degree), urbanity-rurality level (NCHS classification), and age (median age). A spearman correlation provides an opportunity to investigate the monotonic relationship between two continuous variables of demographic features and model accuracy. A monotonic relationship occurs when the variables change together, but not necessarily at the same rate [45,46]. Using the P-value to evaluate the correlation analysis significance, we can assess whether performance is similar (fair) or not (unfair) between social groups.

## RESULTS

**Model Performance**

First, we discuss the COVID-19 case prediction performance of the two models presented (see Figure 2).

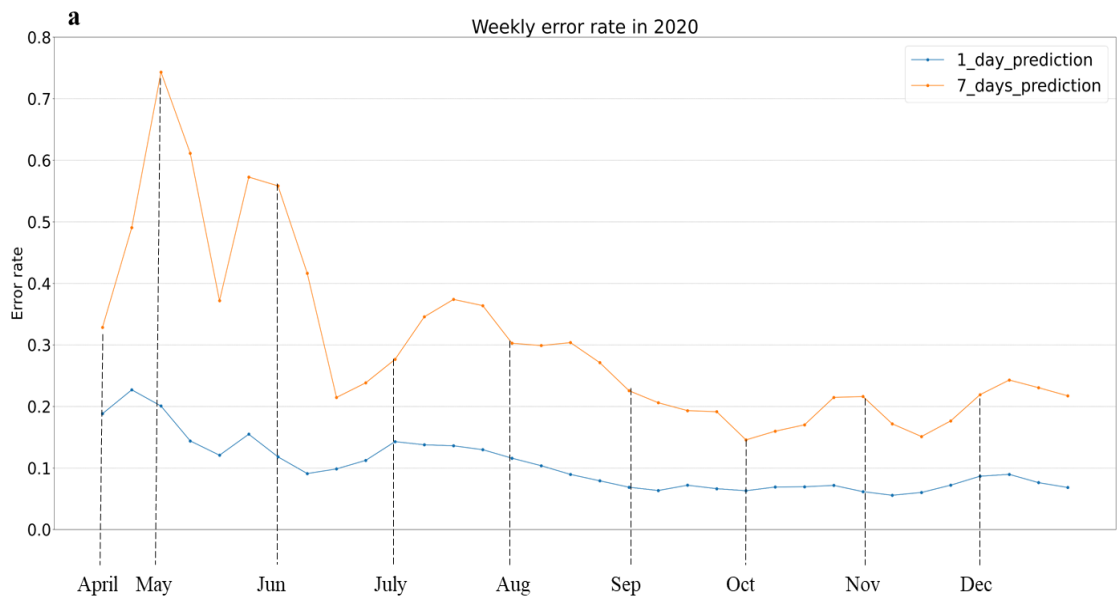

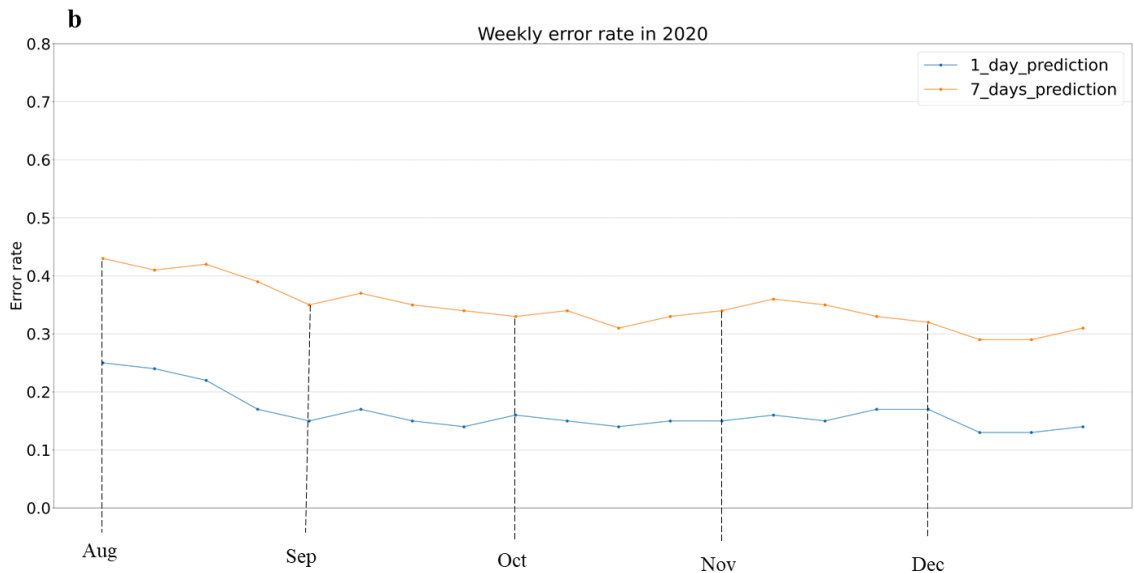

Figure 2. Prediction error rate on a weekly basis. (a) Regression model (b) Time series model

As we can observe, both models predict the number of next day cases (1-day) with an average weekly error rate of 10-20%, and the number of cases in a week (7-days) with an average weekly error rate of 30-40%. Models' performance is in a comparable range to previous studies [21,28,29,41-44], but with the difference that we reported the results for the entire year of 2020 and the US, not specific regions or COVID waves.

**Fairness Performance**

For Model 1, we observe a negative and statistically significant spearman rank correlation between the prediction error rate and income (R (1-day) = -0.13, R (7-days) = -0.08, p-value < 0.001), smartphone ownership (R (1-day) = -0.14, R (7-days) = -0.09, p-value < 0.001), population (R (1-day) = -0.11, R (7-days) = -0.07, p-value < 0.001), bachelor degree (R (1-day) = -0.13, R (7-days) = -0.09, p-value < 0.001). The results suggest that Model 1 – a regression model of COVID-19 cases with mobility – performs better (has fewer errors) in counties with higher incomes, higher smartphone ownership, larger populations, and higher educational levels. On the other hand, correlation analysis indicates a moderate and positive relationship between NCHS code and error rate (R (1-day) = 0.21, R (7-days) = 0.15, p-value < 0.001) and median age (R (1-day) = 0.12, R (7-days) = 0.09, p-value < 0.01). Therefore, as rurality, and age increased, the model's error rate increased, suggesting it performs worse in rural areas and among older and/or black communities (Figure 3 represents the weekly correlations for some of these features).

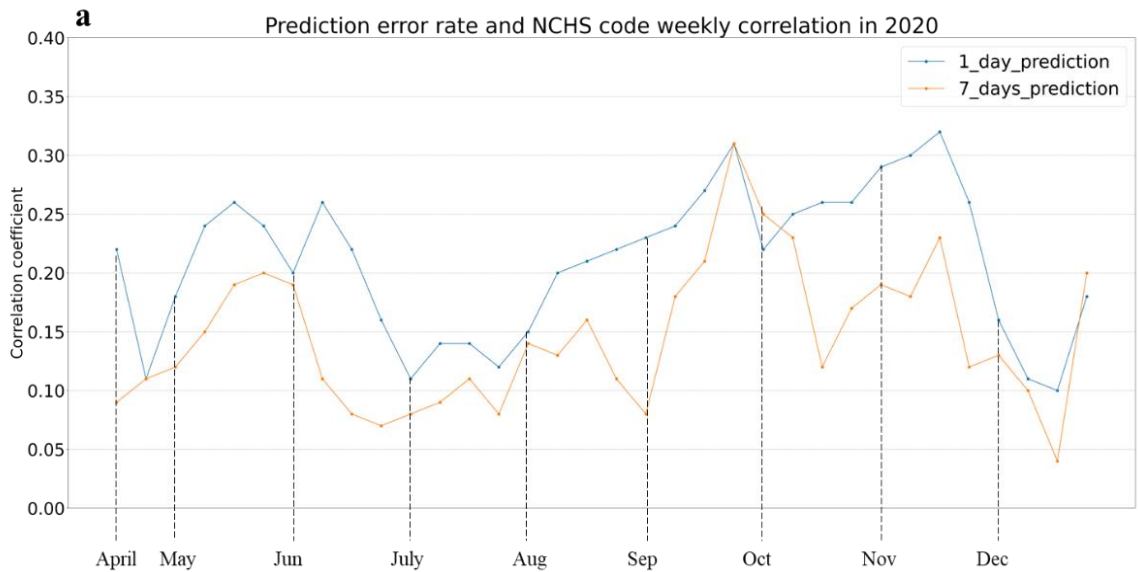

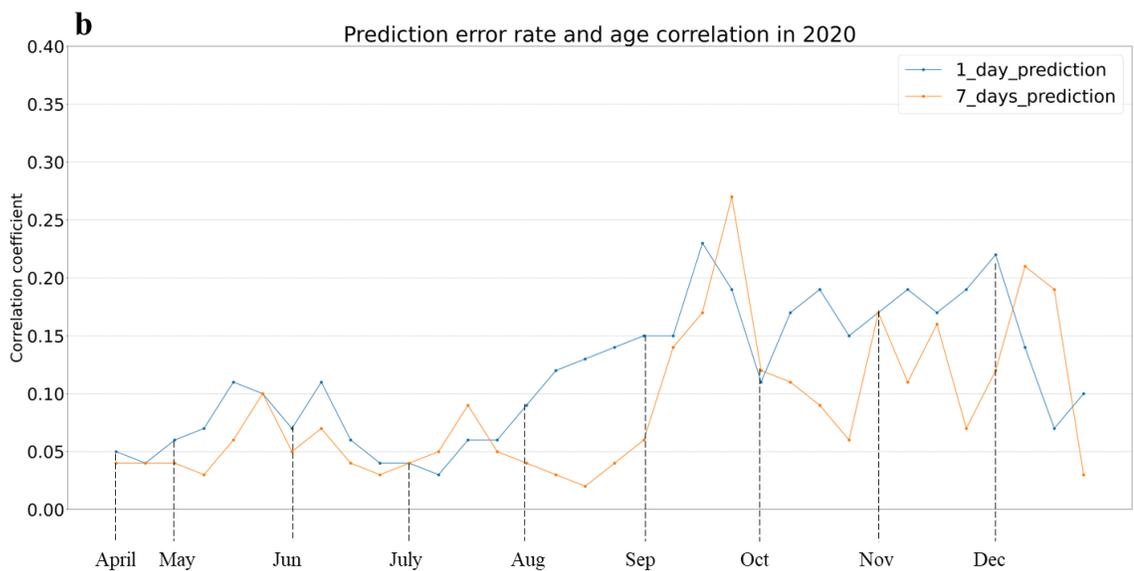

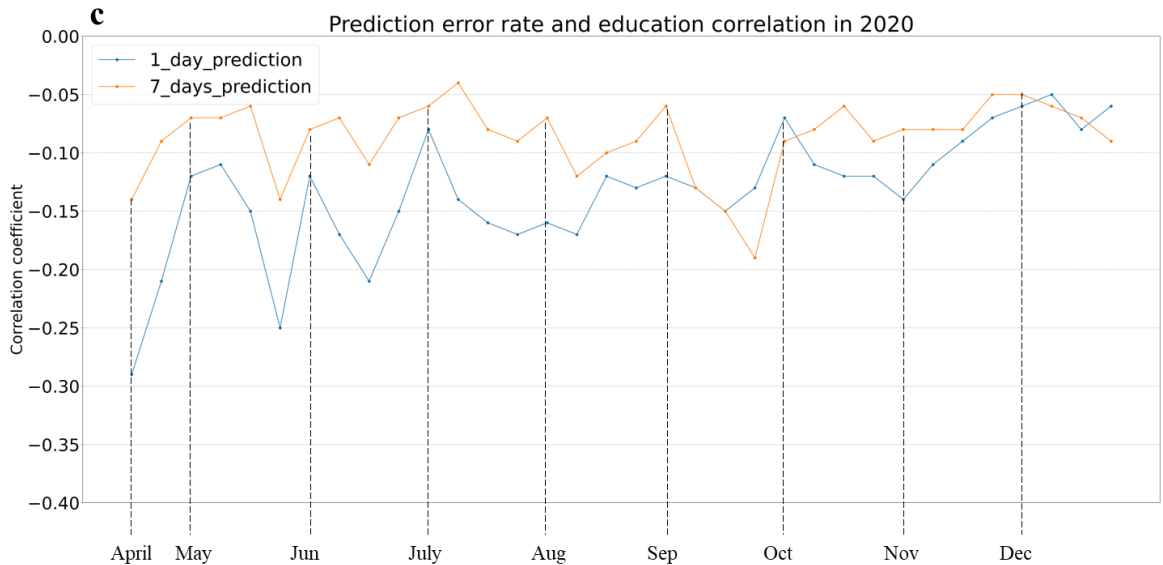

Figure 3. Correlation analysis with selected factors on a weekly basis for Model 1. (a) NCHS code (b) Age, and (c) education.

For Model 2, we observe a negative and statistically significant spearman rank correlation between the prediction error rate and the income (R (1-day) = -0.13, R (7-days) = -0.09, p-value < 0.001), smartphone ownership (R (1-day) = -0.13, R (7-days) = -0.10, p-value < 0.001), population (R (1-day) = -0.11, R (7-days) = -0.11, p-value < 0.001), bachelor degree (R (1-day) = -0.13, R (7-days) = -0.08, p-value < 0.001). These results reveal that Model 2 – an ARIMAX with mobility data added as exogenous variable – performs better (i.e., has lower errors) in counties that share higher income, higher smartphone ownership, larger populations, and higher educational levels (see Figure 4 for weekly representations of the weekly correlations for some of these features). On the other hand, the correlation analysis also reveals a moderate and positive relationship between the error rate and the NCHS code (R (1-day) = 0.20, R (7-days) = 0.21, p-value < 0.001) and median age (R (1-day) = 0.08, R (7-days) = 0.09, p-value < 0.01). In other words, the model's error rate increased as rurality and age increased, revealing a model that performs worse in rural environments, and among older populations. Given that these findings are replicated across models 1 and 2, thus controlling for algorithmic bias, we posit that the model is unfair in part due to the bias in the mobility data used in the model, although bias in the way COVID-19 case data is collected, could also influence the outcome.

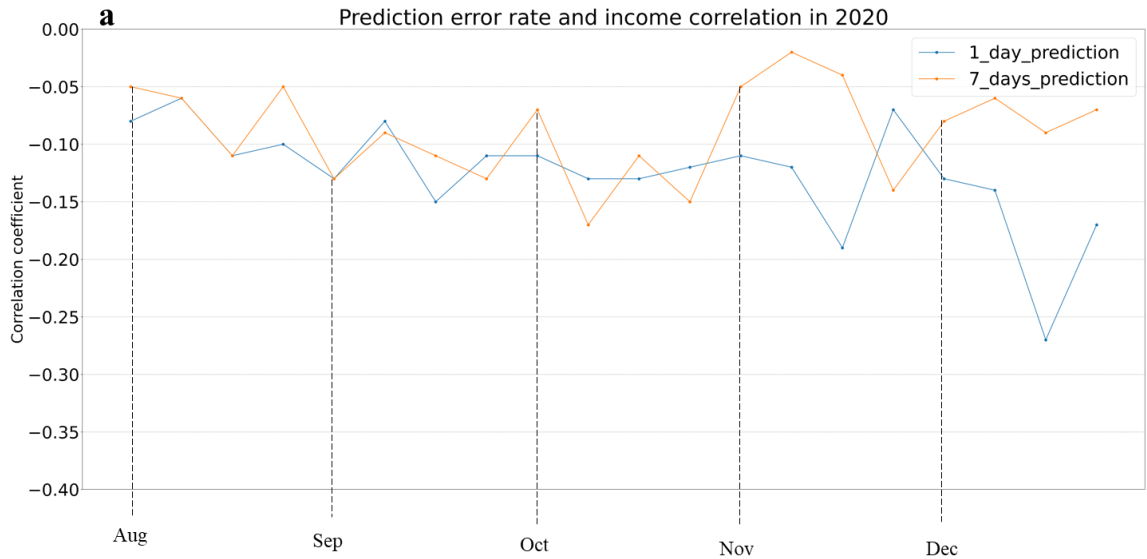

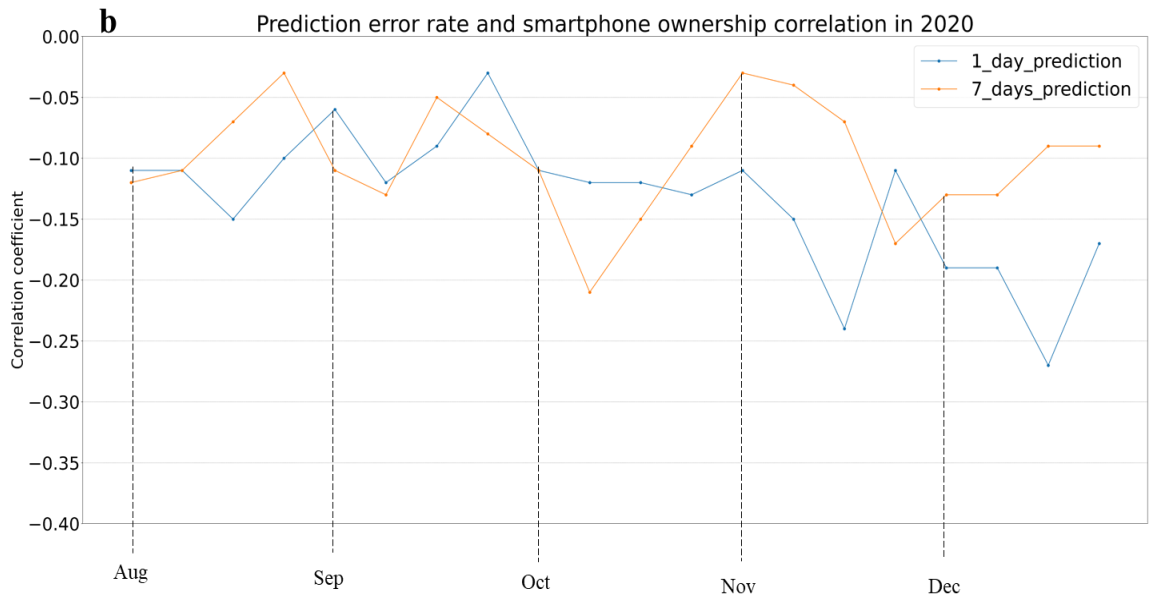

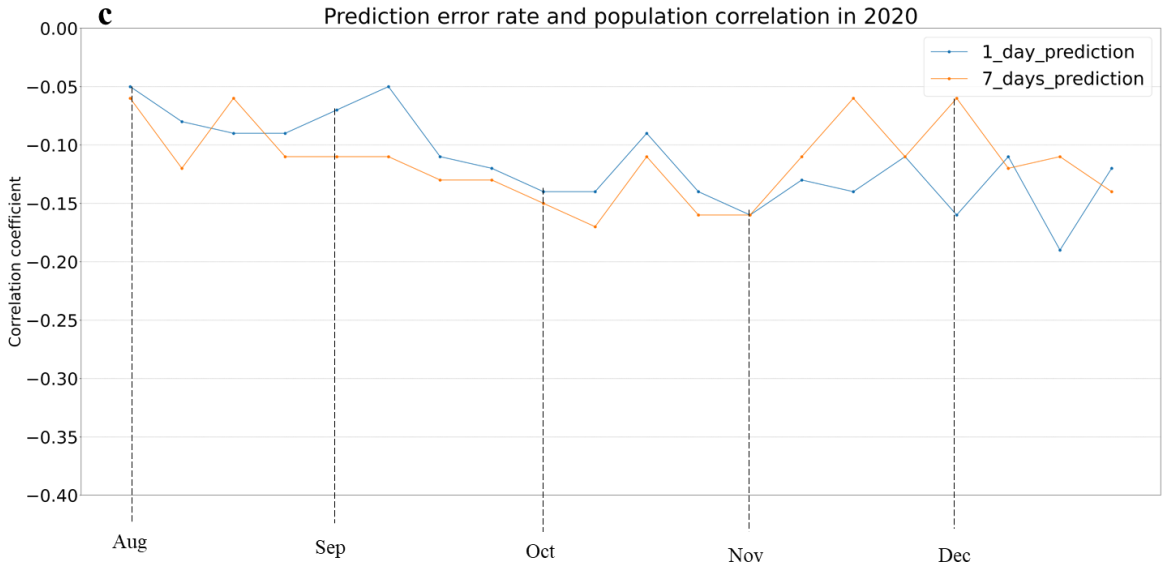

Figure 4. Correlation analysis with selected factors on a weekly basis for Model 2. (a) Income (b) Smartphone ownership, and (c) Population.

To summarize the fairness analysis across models, Tables 1 and 2 provide the monthly correlation averages between the sociodemographic factors at the county level and the error rates for Models 1 and 2 1-day and 7-day predictions, respectively. As discussed, due to the diverse size of the optimal training windows, Model 1 predictions run from April till December, while Model 2 predictions are produced from August till December. With a few fluctuations, and as discussed with in the weekly analyses in Figures 3 and 4, both models show the same pattern of results throughout 2020: lower prediction errors in large, highly educated, wealthy, young, and urban counties.

|  | Income | | Smartphone | | Population | | Education | | NCHS | | Age | |
| --- | --- | --- | --- | --- | --- | --- | --- | --- | --- | --- | --- | --- |
|  | 1 day | 7 day | 1 day | 7 day | 1 day | 7 day | 1 day | 7 day | 1 day | 7 day | 1 day | 7 day |
| **April** | -0.20*** | -0.09*** | -0.19*** | -0.09*** | -0.14*** | -0.07*** | -0.25*** | -0.12*** | 0.17*** | 0.11*** | 0.05* | 0.04* |
| **May** | -0.14*** | -0.06*** | -0.16*** | -0.07*** | -0.14*** | -0.09*** | -0.16*** | -0.09*** | 0.23*** | 0.17*** | 0.08*** | 0.03** |
| **Jun** | -0.14*** | -0.09*** | -0.14*** | -0.07*** | -0.09*** | -0.05*** | -0.16*** | -0.08*** | 0.21*** | 0.11*** | 0.07*** | 0.05*** |
| **July** | -0.14*** | -0.07*** | -0.08*** | -0.06*** | -0.07*** | -0.05*** | -0.14*** | -0.07*** | 0.13*** | 0.09*** | 0.05* | 0.06*** |
| **August** | -0.15*** | -0.07*** | -0.15*** | -0.08*** | -0.11*** | -0.05*** | -0.15*** | -0.10*** | 0.20*** | 0.14*** | 0.12*** | 0.03* |
| **September** | -0.12*** | -0.12*** | -0.19*** | -0.17*** | -0.13*** | -0.10*** | -0.13*** | -0.13*** | 0.26*** | 0.20*** | 0.18*** | 0.16*** |
| **October** | -0.10*** | -0.11*** | -0.16*** | -0.14*** | -0.14*** | -0.11*** | -0.11*** | -0.10*** | 0.26*** | 0.22*** | 0.16*** | 0.13*** |
| **November** | -0.11*** | -0.07*** | -0.17*** | -0.09*** | -0.11*** | -0.07*** | -0.10*** | -0.07*** | 0.29*** | 0.18*** | 0.18*** | 0.13*** |
| **December** | -0.09*** | -0.06*** | -0.09*** | -0.09*** | -0.07*** | -0.06*** | -0.06*** | -0.07*** | 0.14*** | 0.12*** | 0.13*** | 0.14*** |

Table 1. County-level correlations between model 1 error rate and sociodemographic features. (Note: Statistical significance: *** p_value < 0.001, ** p_value < 0.01, * p_value < 0.05)

|  | Income | | Smartphone | | Population | | Education | | NCHS | | Age | |
| --- | --- | --- | --- | --- | --- | --- | --- | --- | --- | --- | --- | --- |
|  | 1 day | 7 day | 1 day | 7 day | 1 day | 7 day | 1 day | 7 day | 1 day | 7 day | 1 day | 7 day |
| **August** | -0.09* | -0.07* | -0.12** | -0.08* | -0.08* | -0.09** | -0.10** | -0.06 | 0.14** | 0.15*** | 0.08* | 0.08** |
| **September** | -0.12*** | -0.12** | -0.08* | -0.09* | -0.09** | -0.12** | -0.14** | -0.09** | 0.15** | 0.16*** | 0.05* | 0.05* |
| **October** | -0.12** | -0.13*** | -0.12*** | -0.14*** | -0.13** | -0.15*** | -0.11*** | -0.09** | 0.19*** | 0.23*** | 0.08* | 0.08** |
| **November** | -0.12*** | -0.06* | -0.15*** | -0.08* | -0.14*** | -0.11*** | -0.14*** | -0.07* | 0.26*** | 0.31*** | 0.11* | 0.16** |
| **December** | -0.18*** | -0.08** | -0.21*** | -0.11** | -0.15*** | -0.11*** | -0.17*** | -0.08*** | 0.27*** | 0.22*** | 0.09** | 0.09** |

Table 2. County-level correlations between model 2 error rate and sociodemographic features. (Note: Statistical significance: *** p_value < 0.001, ** p_value < 0.01, * p_value < 0.05)

# DISCUSSION

To combat the COVID-19 pandemic, governments and private companies around the world were promoting the use of digital public health technologies for data collection and processing [47-50]. Through the use of GPS, cellular networks and Wi-Fi, smartphones can collect and aggregate location data in real-time to monitor population flows, identify transmission hotspots, determine the effectiveness of NPIs [51], and predict future COVID-19 cases [7,20,25].

Using SafeGraph's mobility data, we examined whether two popular predictive models that use mobility data to predict COVID-19 cases over time, performed fairly across social groups. Our findings revealed a correlation between a county's socio-economic and demographic characteristics and the models' error rates. In particular, we observed that the prediction errors were lower in large, highly educated, wealthy, young, and urban counties. Given that the findings were similar across models, thus controlling for algorithmic bias, we posit that the presence of bias in the mobility data negatively impacts the model predictions by unfairly outputting case numbers with higher errors for specific social groups. Furthermore, our results show that mobility data appears to be less likely to capture older, poorer, non-white, and less educated users. Thus, allocating public health resources based on such mobility data could disproportionately harm seniors and minorities at high risk. We acknowledge that these unfair results could also be in part due to bias in COVID-19 data collection processes, which might suffer from higher under-reporting by certain social groups, and our future work will look into this.

To generalize smartphone-derived insights over a population, the mobility data must reveal information about the population without bias i.e., information that is representative across socio-economic and demographic groups. However, due to the lack of ground truth data about the socio-economic and demographic characteristics of the population whose mobility data is collected, this study has also shown that investigating performance fairness can provide valuable insights into potential mobility data biases.

Finally, as the research community moves forward with the use of mobility data in COVID-19 case prediction models, we think it is important to consider the following set of recommendations. First, and whenever possible, we strongly suggest applying sampling bias mitigation approaches to correct for under-represented groups in the data. Second, mitigation approaches might not always be possible, due to the lack of demographic information about the individuals whose cell phone data is being collected. For that reason, we encourage the research community working with mobility data to report fairness analyses together with the performance of the predictive models proposed. We hope that these practices will enhance pandemic management via case prediction models that are more transparent and fairer, and that will allow for more equitable decision making.

## Data availability

All data we use in this study are all publicly available. These datasets include the SafeGraph mobility data (. https://docs.safegraph.com/docs/social-distancing-metrics), COVID-19 confirmed cases (https://github.com/CSSEGISandData/COVID-19), and sociodemographic information at US counties level (https://www.census.gov/data/datasets.html).

## Author contributions

The authors confirm contribution to the paper as follows: study conception and design: V.F and A.E; data collection: A.E; analysis and interpretation of results: A.E and V.F; draft manuscript preparation: A.E and V. F. Both authors reviewed and approved the final version of the manuscript.

## Availability of Data and Materials

Correspondence and requests for materials should be addressed to A.E.